\documentstyle[prl,aps,amsfonts,amssymb,floats,epsfig]{revtex}
\begin{document}
\draft 
\twocolumn[\hsize\textwidth\columnwidth\hsize\csname @twocolumnfalse\endcsname
\title{Mid-infrared absorption in YBa$_2$Cu$_3$O$_6$: \\ 
Failure of spin-wave theory in undoped cuprates? }
\author{M. Gr\"uninger$^1$, D. van der Marel$^1$, A. Damascelli$^1$, A. Erb$^2$,  
Th.\ Wolf$^3$, T. Nunner$^4$, and T. Kopp$^4$}
\address{$^1$\it Lab.\ of Solid State Physics, MSC, Univ.\ of Groningen, The Netherlands;
{\rm $^2$} DPMC, Univ.\ of Geneva, Switzerland;\\
\hskip 2mm {\rm $^3$} ITP, Forschungszentrum Karlsruhe, Germany;
{\rm $^4$} TKM, Univ.\ of Karlsruhe, Germany}
\date{March 17, 1999}
\maketitle
\begin{abstract}
The optical conductivity $\sigma (\omega )$ of undoped YBa$_2$Cu$_3$O$_{6}$ 
is studied in detail in the mid-infrared range. Substitutions on all but the Ba 
site are used to identify the prominent absorption processes at 2800 and 3800 cm$^{-1}$. 
Experimental evidence for bimagnon--plus--phonon absorption is collected. 
A more critical analysis of the lineshape and the spectral weight reveals the limits 
of this approach. Although phonon-2-magnon multiple scattering seems to reproduce the 
lineshape, the necessary coupling is unrealistically large. 
The strong increase of high frequency spectral weight with increasing temperature makes 
the failure of spin-wave theory even more evident.
\end{abstract}
\pacs{PACS numbers: 74.72.-h 74.25.Gz, 78.30.Hv, 75.40.Gb, 75.50.Ee}
]
\narrowtext
The undoped parent compounds of the high T$_c$ cuprates are regarded as an almost 
ideal realization of a two-dimensional (2D) spin 1/2 Heisenberg antiferromagnet. 
Despite the low dimensionality and the low spin the excitations are thought to be 
spin-waves with a well-defined dispersion\cite{manousakis}, as opposed to e.g.\ 1D 
systems, where a spinon continuum is observed in neutron scattering\cite{1D}.
In the cuprates, a spin-wave dispersion has been extracted throughout the whole 
Brillouin zone from the maxima in neutron scattering intensities, but energies are 
rather high, large backgrounds are observed and the magnitude of quantum 
corrections is unclear\cite{bourges97}.
Moreover, the assumption that magnons are {\it not} well-defined particles at the Brillouin 
zone boundary was a keypoint in the successful description of the photoemission data 
of insulating Sr$_2$CuO$_2$Cl$_2$ in Ref.\ \cite{chubukovPE}. 
Two-magnon (2M) Raman scattering shows several anomalies 
in the cuprates, in particular a very broad lineshape, spectral weight at high energies 
and a finite signal in $A_{\it 1g}$ geometry\cite{blumberg}. 
A large body of theoretical work has been dedicated to this problem, and the importance 
of resonance phenomena\cite{chubukov} was emphasized. Other treatments include the 
interaction with phonons\cite{nori} and extensions of the Heisenberg model\cite{eroles}. 
Certainly, the strong influence of the charge transfer (CT) resonance on the Raman spectra 
complicates the problem significantly. 

Optical spectroscopy probes the magnetic excitations more directly. 
The main peak in mid-infrared absorption (MIR) spectra of 
La$_2$CuO$_4$ and other single layer cuprates\cite{perkins} has been 
interpreted by Lorenzana and Sawatzky \cite{lorenzana} in terms of bimagnon--plus--phonon 
(BIMP) absorption. A similar feature was reported in the bilayer system  
YBa$_2$Cu$_3$O$_6$ (YBCO$_6$)\cite{grueninger}, 
in 2D {\bf S}=1 La$_2$NiO$_4$\cite{lorenzana,perkinsNi} and 
in 1D {\bf S}=1/2 Sr$_2$CuO$_3$\cite{suzuura,lorenzanaeder}. 
Good agreement is achieved in 1D \cite{lorenzanaeder} because quantum fluctuations are 
included ab initio, and for the 2D {\bf S}=1 nickelates \cite{lorenzana} because fluctuations 
beyond spin-wave theory are small. 
However, in the cuprates the estimate for the spectral weight of BIMPs is one order of magnitude 
too small. Moreover, a large amount of spectral weight is observed above the BIMP absorption. 
Interpretations in terms of multi-magnon--plus--phonon absorption\cite{lorenzana}, 
$d$-$d$ transitions\cite{perkinsNi} and CT excitons\cite{wang} have been proposed. 
We challenge these approaches and suggest that a full account of our MIR data in the undoped 
cuprates has to include quantum fluctuations beyond spin-wave theory. This might provide an 
important feedback to the Raman experiment and the basic picture of the undoped cuprates. 

Specific ionic substitutions help to identify the prominent resonances in the MIR spectrum 
of YBCO$_6$. In particular we use oxygen isotope substitution to distinguish vibrational 
from electronic degrees of freedom and to give direct experimental evidence for the existence 
of BIMP absorption. 
Substituting rare earth (RE) elements for Y increases the lattice parameter $a$ and 
thereby changes the exchange constant $J$. 
The dependence of the BIMP frequency on $J(a)$ is found to be similar to the one 
reported for 2M Raman scattering\cite{yoshida,cooper}.

Single crystals of YBa$_2$Cu$_3$O$_7$ were grown in BaZrO$_3$ (BZO) crucibles. 
Crystals grown using this technique exhibit a superior 
purity ($>$ 99.995 at.\ \%)\cite{erb1,erb2}.
The full exchange of the oxygen isotope was described elsewhere\cite{crete}.
For comparison, single crystals grown in Y$_2$O$_3$ stabilized ZrO$_2$ (YSZ) 
crucibles\cite{erbysz} were studied as well. 
A finite amount of Y in the single crystals of RE$_{0.8}$Y$_{0.2}$Ba$_2$Cu$_3$O$_6$ 
(RE=Pr, Gd) studied here is due to the Y$_2$O$_3$ in the ZrO$_2$ crucibles.
In the case of YBa$_2$Cu$_{3-y}$Zn$_y$O$_6$\cite{wolf}, we 
estimate $0.05 \! \le \! y \! \le 0.07$\cite{crete}. 
The oxygen content of all crystals was fixed to a value O6 by annealing them for about 
a week in a flow of high purity Argon (99.998 \%) at 750$^\circ $C or 
in ultra high vacuum at 700$^\circ $C.
We calculated $\sigma (\omega)$ by inverting the Fresnel equations for the experimentally 
measured transmission and reflection data. Measurements were carried out with the 
electric field vector polarized parallel and perpendicular to the $ab$-plane. 
The small remnants of interference fringes in some of the calculated spectra 
of $\sigma (\omega)$ are artifacts caused by deviations of the measured data 
from the assumed ideal case of absolutely flat and plane parallel surfaces.

\begin{figure}[t]
\centerline{\psfig{figure=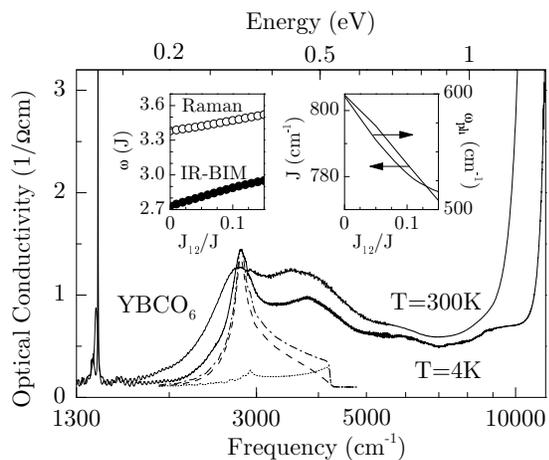,width=7.2cm,clip=}}
  \caption{$\sigma(\omega)$ of YBCO$_6$ at 4 and 300 K.\@ Dashed and dotted lines: In- 
  and inter-plane contributions to BIMP absorption for $J\!=\!780$ cm$^{-1}$, $J_{12}/J\!=\!0.1$ 
  and $\hbar\omega_{ph}\!=\!530$ cm$^{-1}$. Dash-dotted line: sum of the two. 
  Left inset: 2M Raman and IR BIM frequencies as a function of $J_{12}/J$. Right inset: 
  $J$ and $\hbar\omega_{ph}$ as a function of $J_{12}/J$. }
\end{figure}

\begin{figure}[t]
\centerline{\psfig{figure=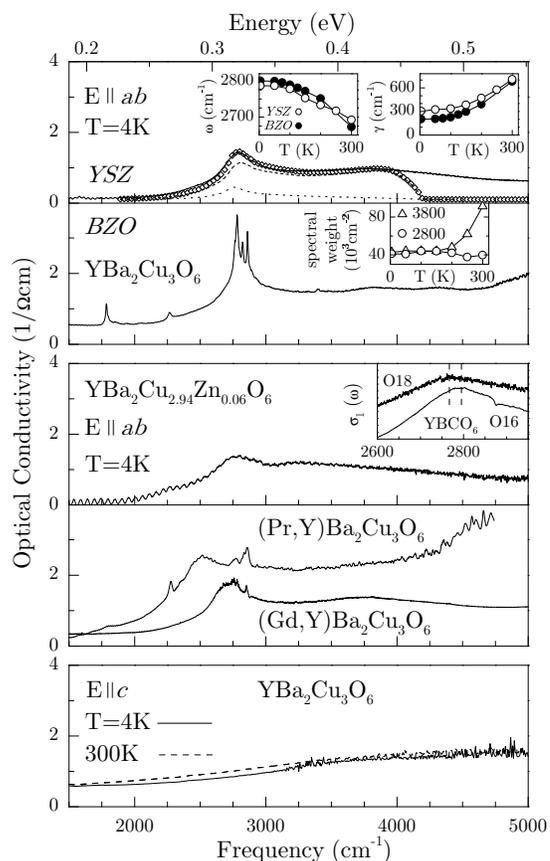,width=7.2cm,clip=}}
  \caption{$\sigma (\omega )$ for different substitutions and polarizations. 
  Diamonds and broken lines in the top panel are theoretical results. See text 
  for details. Insets top panel: T dependence of peak frequency and width of the 
  2800 cm$^{-1}$ peak in  YBCO$_6$ samples grown in YSZ (open circles, top panel) and 
  BZO (full circles, second panel). 
  Inset panel 2: Spectral weight of the 2800 and 3800 cm$^{-1}$ peaks. }
\end{figure}

We first collect evidence for the applicability of a magnon-phonon interpretation 
of the main resonance. In Fig.\ 1 we display $\sigma (\omega)$ of YBCO$_6$ 
up to the onset of CT absorption for T=4 and 300 K.\@ 
Note the very low values of $\sigma (\omega)$, which are two (four) orders 
of magnitude lower than for YBCO$_{6.1}$ (YBCO$_7$) in this frequency range. 
Following the interpretation of the single layer compounds\cite{lorenzana} we ascribe 
the main peak at 2800 cm$^{-1}$ to BIMP absorption. A magnetic origin is also favored 
by recent measurements of the pressure dependence of the MIR and Raman spectrum of 
Sr$_2$CuO$_2$Cl$_2$\cite{graybeal}.
In the cuprates bimagnon absorption is forbidden due to inversion symmetry. It only 
becomes weakly allowed by symmetry breaking effects like impurities or the combination 
with a phonon. Let us compare the experimental data with predictions of BIMP theory on a 
qualitative basis (Fig.\ 2). The two upper plots show spectra of two samples of 
YBCO$_6$ grown in different crucibles (top panel YSZ, below BZO). 
The strong temperature dependence of the BIMP peak frequency is most likely due to the reduction 
of spin stiffness with increasing temperature (left inset, open/full symbols YSZ/BZO). 
The width $\gamma $ of the BIMP resonance is smaller in the cleaner sample grown in 
a BZO crucible (right inset)\cite{para}. 
We therefore expect that impurity scattering is not negligible in the 
determination of the correct lineshapes. 
Substituting Zn on Cu sites indeed broadens the 2800 cm$^{-1}$ peak drastically 
(panel 3 from top in Fig.\ 2).
A finite phonon contribution to the BIMP peak is evident from the frequency shift induced by 
oxygen isotope substitution in YBCO$_6$ (see inset of same panel). The measured isotope shift 
of $28 \pm 8$ cm$^{-1}$ is consistent with the BIMP interpretation, assuming that the 
longitudinal stretching phonon of approximately 550 -- 600 cm$^{-1}$ is excited.
Substitution of Y with Pr or Gd leads to a significant frequency shift of the main peak. 
Similar shifts were observed in 2M Raman scattering\cite{yoshida,cooper} and were explained 
by the dependence of $J$ on the lattice parameter $a$. 
Similar to the case of Zn, the disorder on the Y site after substitution of 80\% 
of Pr or Gd enhances $\gamma $. Finally, the 2800 cm$^{-1}$ 
peak is not observed if the electric field is applied parallel to the $c$ axis, in 
agreement with the single layer data\cite{perkins}. 
The spikes on top of the main resonance and the other sharp features have not yet 
been identified. A relation to BIMP absorption is unlikely, as their position 
is identical in the RE samples. 

For a more critical analysis of peak frequency, lineshape and spectral weight 
we extend the BIMP theory to the bilayer case. In order to obtain the coupling to 
light we start from a Heisenberg Hamiltonian which takes into account a 
dependence of the in- and inter-plane exchange constants $J$ and $J_{12}$ 
on the external electric field {\bf E} and the phonon coordinates\cite{lorenzana}:
\begin{displaymath}
H  =  \sum_{L=1,2}\sum_{<i,j>}J({\bf E},{\bf u})
{\bf S}_{L,i}{\bf S}_{L,j} 
 +  \sum_{i}J_{12}({\bf E},{\bf u}){\bf S}_{1,i}{\bf S}_{2,i}  \nonumber
\label{1}
\end{displaymath}
where $i$ and $j$ label nearest neighbor Cu sites in a 2D square lattice, $L$ labels the 
two planes in a single bilayer, and ${\bf u}$ denotes the displacements of 
O ions. Only Einstein phonons are considered. The different phonons modulate the 
intersite hopping and the on-site energies on both Cu and O sites. Modulations are taken 
into account to second order since we also included phonon--2M {\it multiple} scattering  
processes in a refined approach (see below). In Ref.\ \cite{lorenzana} a high energy (HE) 
approximation was used which is inappropriate for zone center excitations. 
Our RPA results for two interacting magnons show that the HE approximation reproduces 
the line shape rather well. Both RPA and HE produce 2M bound states with a strong 
dispersion in momentum space. However, the sharp resonances at the low energy side of 
the BIMP reported in Ref.\ \cite{lorenzana} are removed in RPA.

Both the Raman and infrared 2M peak frequencies were calculated as a function of 
$j\!=\!J_{12}/J$ (left inset of Fig.\ 1). In the infrared case, the phonon frequency 
$\hbar\omega_{ph}$ still has to be added. 
At T=4K the experimental BIMP and 2M Raman spectra peak at 2795 and 
2720 $\pm 10$ cm$^{-1}$\cite{blumberg}, respectively. From these we can deduce 
the values of $J$ and $\hbar\omega _{ph}$ for a given ratio $J_{12}/J$ (right inset of 
Fig.\ 1). We obtain $J\!=\!790\!\pm\!10$ cm$^{-1}$ and $j\!=\!0.08\!\pm\!0.04$. 
Neutron scattering suggests $j$=0.1 -- 0.15\cite{reznik,hayden} and 
$\hbar\omega _{ph}\approx 550$ -- 600 cm$^{-1}$\cite{reichardt} for the relevant longitudinal 
stretching phonon mode.
However, a finite next-nearest neighbor coupling $J'$ will shift the values of $J$ and
$J_{12}$ considerably\cite{morr}.

Let us compare the calculated and measured lineshapes and oscillator strengths of the 
BIMP peak. The calculated BIMP absorption 
for $J$=780 cm$^{-1}$, $J_{12}$=0.1$J$ and $\hbar\omega _{ph}\!=\!530$ cm$^{-1}$ is plotted 
together with the experimental curve in Fig.\ 1 (dash-dotted line). 
An offset of 0.1 $\Omega ^{-1}$cm$^{-1}$ has been used. In a bilayer we have to distinguish 
two contributions: a photon can flip two spins (i) in the 
same layer (dashed line, in-plane) or (ii) in adjacent layers (dotted line, inter-plane). 
For (i) a rough estimate of the 2M binding energy in the Ising limit is $J$. 
This implies that the 2M energy at the zone boundary is about $J$ below the upper 
cut-off of the 2M-spectrum, which approximately corresponds to the BIMP position in Fig.\ 1. 
Similarly, the binding energy is $J_{12}$ for case (ii) which explains the maximum of the 
calculated inter-plane conductivity just below the 2M cut-off.  
The estimated relative spectral weight of 
inter- and in-plane contributions is 0.06 -- 0.3 for $j$=0.1. Due to this small value the 
spectral weight is similar in YBCO$_6$ and in La$_2$CuO$_4$ both experimentally and 
theoretically. The perturbatively estimated spectral weight is a factor of 8 -- 15 too 
small compared to experiment\cite{qA}. 
\begin{table}[t]
\begin{tabular}{rcccc}
 & peak A & peak B & A/B  & 2M cut-off        \\ \hline
 YBa$_2$Cu$_3$O$_6$ & 2800 & 3800  & 0.74 & 3700  = 4.72$J$   \\
 La$_2$CuO$_4$ & 3300$^*$ & 4500$^*$ & 0.73 & 4500 = 4.63$J$ \\
 Sr$_2$CuO$_2$Cl$_2$ & 2900$^*$ & 4000$^*$ & 0.73 & 4000 = 4.63$J$
\end{tabular}   
\caption{Measured frequencies of the two main MIR absorption peaks, 
their ratio, and the 2M cut-off (all frequencies in cm$^{-1}$).
$^*$ taken from Ref.\ [9]. }
\end{table}

Regarding the lineshape the calculated curve is sharper than the experimental one, 
but the width of the BIMP peak is sample dependent, as stated above. 
As in the single layer cuprates, the real problem is obviously at higher frequencies: 
the strong peak at 3800 cm$^{-1}$ remains unexplained. It is likely that the high 
energy anomaly has the same origin in MIR and Raman spectra. This is substantiated 
by the absence of the anomaly in both spectroscopies in {\bf S}=1 
La$_2$NiO$_4$\cite{perkinsNi,sugai}. In several cuprates, the frequency ratio of the two 
dominant MIR peaks is about 0.73 (Table 1), strongly suggesting a common magnetic origin.
In other terms, the second peak in both single and bilayers is close to the 2M cut-off. 
One way of shifting spectral weight to the 2M cut-off is to consider a finite interaction 
between phonons and bimagnons. A dimensionless coupling constant for phonon-2M multiple   
scattering is defined as 
$
\lambda _{p2M}\!=\!\frac{1}{2J}\langle \frac{d^2J}{du^2} \rangle \langle u^2 \rangle , 
$
from which we estimate $\lambda _{p2M} \approx -0.02...+0.01$ for the stretching phonon 
mode\cite{harrison,lambda}. A negative value of $\lambda _{p2M}$ translates into a 
repulsive phonon-2M interaction and shifts spectral weight to higher frequencies. 
A way to test the reliability of our estimate of $\lambda _{p2M}$ is to compare the 
{\it linear} coupling $dJ/du$ with the experimental pressure dependence 
of $J$\cite{aronson}. There, our estimate is 1 -- 2 times smaller. 
However, an excellent fit to the data (diamonds in top panel of Fig.\ 2) is obtained 
only if we assume $\lambda _{p2M}=-0.2$ (dashed line) and add the BIMP contribution of 
the apical stretching phonon. The apical contribution is expected to have a 5 times smaller 
weight and a negligible phonon-2M coupling (dotted line). This large value of 
$\lambda _{p2M}$ makes such a scenario very unlikely.
Phonon-magnon scattering processes have also been considered for the explanation 
of the width of the 2M Raman resonance\cite{nori,eroles}. 
Contrary to our dynamic treatment, their adiabatic approach models static 
disorder, which enhances the width but does not result in a second 
resonance. Hence we conclude that phonon-magnon interaction processes 
cannot explain the magnetic MIR and Raman anomalies.

A severe constraint for any interpretation of the high frequency spectral weight 
is the observed increase by a factor of more than 2 from 4 to 300 K (triangles in 
inset of panel 2 of Fig.\ 2). A similar behavior can be detected in the temperature 
dependence of $\sigma (\omega )$ of Sr$_2$CuO$_2$Cl$_2$\cite{ziboldSCOC}. 
We propose that these findings support the notion of a strong local deviation 
from the N{\'e}el state which is even more pronounced for 300 K.\@ Whereas the broken 
symmetry of the antiferromagnetic state will still support long wavelength 
spin-wave excitations, the character of the short wavelength magnetic 
excitations reflects the strong quantum fluctuations and consequently they are 
insufficiently represented by spin-waves. 
Both Raman and MIR are dominated by short wavelength magnetic excitations which makes 
their evaluation within spin-wave theory less reliable. 
We emphasize that this interpretation does not contradict the good agreement of neutron 
scattering results with spin-wave theory for small momenta. 
Note that even in 1D the inapplicability of a spin-wave picture to neutron data was not realized 
for many years\cite{endoh}.

Exact diagonalization should serve to identify the weight of magnetic excitations.
However, only clusters of up to $\sqrt{20} \times \sqrt{20}$ were investigated and produced 
a minor contribution to MIR absorption at high frequencies\cite{sawatzky}. 
Since this cluster size is still comparable to the size of the considered high frequency  
excitations, a finite size scaling analysis would be a serious test whether the Heisenberg 
model or a 4-spin extension\cite{eroles} of it can generate the measured high frequency weight.

In the absence of doping, the only alternative to a magnetic origin of the anomaly are excitons. 
Perkins {\em et al.}\cite{perkinsNi} suggested a $d$-$d$ exciton with sidebands. However, 
this exciton should not be IR active, 
and there is theoretical\cite{eskes,martin} and experimental\cite{kuiper} evidence that its 
actual energy is a factor of two to three higher. 
Wang {\it et al.}\cite{wang} predict a CT exciton at 0.8 eV (6500 cm$^{-1}$) from fits to 
EELS data between 2.5 and 4 eV.\@ We consider the Coulomb attraction necessary to pull  
this exciton down to 3800 cm$^{-1}$ -- far below the CT gap -- as 
unrealistically large. Furthermore, a CT exciton should follow the strong redshift of the 
onset of CT absorption with increasing temperature (10500 to 9000 cm$^{-1}$, see Fig.\ 1), 
which is not observed. Both exciton models fail to describe the increase of spectral 
weight with temperature. 
Hence the exciton interpretation is an unlikely scenario for the considered resonance. 
Direct excitation of 2 magnons in bilayers via spin-orbit 
coupling was suggested\cite{grueninger} to reproduce the MIR lineshape in YBCO$_6$ up to 
4000 cm$^{-1}$. However, a spin exchange of $J_{12}=0.5 J$ had to be assumed, which is 
not anymore consistent with neutron scattering\cite{reznik,hayden}.

We conclude that present day understanding of magnetic excitations in undoped 
cuprates is not sufficient to explain both MIR and Raman data. Only a more appropriate 
treatment of the short wavelength excitations will explain the observed anomalies.

It is a pleasure to acknowledge many stimulating discussions with G.A. Sawatzky. 
We also want to thank R. Eder and J. Brinckmann for helpful discussions. 
This project is supported by the Netherlands Foundation for Fundamental Research 
on Matter with financial aid from the Nederlandse Organisatie voor 
Wetenschappelijk Onderzoek and by the DFG and DFG-GK.


\begin{references}
\bibitem{para}The parameters were obtained from a Lorentzian fit to the 2800 cm$^{-1}$ peak. 
The sharp structures of the sample grown in BZO were fit separately.
\bibitem{qA}The deviation is larger than given in Ref.\ \cite{lorenzana}, as a 
realistic estimate of $q_A/e=-4JU_{pd}/\Delta ^2$ (see Ref.\ \cite{lorenzana}) 
yields $q_A/e \approx -0.05$ for both La$_2$CuO$_4$ and YBCO$_6$.
\bibitem{lambda}This is derived from Harrison's 
$t_{pd}(u)\approx t\times(1+u/a_{pd})^{7/2}$\cite{harrison} and modulations of the 
on-site potentials.
\end{references}
\end{document}